\documentclass[10pt, conference, hidelinks]{IEEEtran}
\IEEEoverridecommandlockouts

\RequirePackage{cite}
\RequirePackage{amsmath,amssymb,amsfonts}
\RequirePackage{algorithmic}
\RequirePackage{graphicx}
\RequirePackage{float}
\RequirePackage[caption=false]{subfig}
\RequirePackage{textcomp}
\RequirePackage[dvipsnames]{xcolor}
\RequirePackage{orcidlink}
\RequirePackage[acronym]{glossaries}
\RequirePackage[detect-all=true,per-mode=symbol,per-symbol =/]{siunitx}
\RequirePackage[capitalise]{cleveref}
\RequirePackage{tablefootnote}
\RequirePackage{threeparttable}
\RequirePackage{booktabs}
\RequirePackage{listings}
\RequirePackage{graphicx}
\RequirePackage{microtype}

\DeclareSIUnit{\x}{\texttimes}
\DeclareSIUnit\gateeq{GE}
\ifdefined\review
\newcommand\camr[1]{\textcolor{blue}{#1}}
\else
\newcommand\camr[1]{#1}
\fi

\lstdefinelanguage{other}{
    basicstyle=\itshape,
}

\lstdefinelanguage{abc}{
    language=tcl,
    columns=fullflexible,
    keepspaces=true,
    upquote=true,
    showstringspaces=false,
    basicstyle=\normalsize,
    commentstyle=\color{gray},
    keywordstyle=\color{Blue}\bfseries,
    morekeywords={&get, &st, &if, &syn2, &b, &dch, &put, &nf, rec_add3, rec_start3, read_lib}
}

\lstdefinelanguage{yosys}{
    language=tcl,
    columns=fullflexible,
    keepspaces=true,
    upquote=true,
    showstringspaces=false,
    basicstyle=\normalsize,
    commentstyle=\color{gray},
    keywordstyle=\color{Sepia}\bfseries,
    morekeywords={booth, maccmap, alumacc, extract, abc, abc9}
}

\newacronym{qor}{QoR}{quality of results}
\newacronym{soc}{SoC}{system-on-chip}
\newacronym{rtl}{RTL}{register transfer level}
\newacronym{pnr}{P\&R}{place and route}
\newacronym{eda}{EDA}{electronic design automation}
\newacronym{ast}{AST}{abstract syntax tree}
\newacronym{rtlil}{RTLIL}{RTL intermediate language}
\newacronym{asic}{ASIC}{application-specific integrated circuit}
\newacronym{mac}{MAC}{multiply-accumulate}
\newacronym{fma}{FMA}{fused multiply-add}
\newacronym{drc}{DRC}{design rule check}
\newacronym{nda}{NDA}{non-disclosure agreement}
\newacronym{mpw}{MPW}{multi-project wafer}
\newacronym{csa}{CSA}{carry-save adder}
\newacronym{cpa}{CPA}{carry-propagate adder}
\newacronym{cia}{CIA}{carry-increment adder}
\newacronym{rca}{RCA}{ripple-carry adder}
\newacronym{ppa-bk}{PPA-BK}{Brent-Kung parallel-prefix adder}
\newacronym{ppa-sk}{PPA-SK}{Sklansky parallel-prefix adder}
\newacronym{cmos}{CMOS}{complementary metal-oxide semiconductor}
\newacronym{aig}{AIG}{and-inverter graph}
\newacronym{lms}{LMS}{lazy man's synthesis}
\newacronym{orfs}{ORFS}{OpenROAD flow scripts}
\newacronym{pdk}{PDK}{process design kit}
\newacronym{fsm}{FSM}{finite-state machine}
\newacronym{fpu}{FPU}{floating-point unit}
\newacronym{lec}{LEC}{logic equivalence checking}
\newacronym{hitl}{HITL}{human-in-the-loop}
\newacronym{fpga}{FPGA}{field-programmable gate array}
\newacronym{lut}{LUT}{lookup table}
\newacronym{oseda}{OS EDA}{open-source EDA}
\newacronym{oshw}{OSHW}{open-source hardware}
\newacronym{ip}{IP}{intellectual property}
\newacronym{rot}{RoT}{root of trust}
\newacronym{sv}{SV}{SystemVerilog}
\newacronym{gpos}{GPOS}{general-purpose operating system}
\newacronym{axi}{AXI4}{Advanced eXtensible Interface 4}
\newacronym{llc}{LLC}{last-level cache}
\newacronym{spm}{SPM}{scratchpad memory}
\newacronym{dma}{DMA}{direct memory access}
\newacronym{gpt}{GPT}{GUID partition table}
\newacronym{c2c}{C2C}{chip-to-chip}
\newacronym{at}{AT}{area-time}

\newcommand\riscv{\mbox{RISC-V}}
\newcommand\sv{\mbox{\gls{sv}}}

\title{
Basilisk: An End-to-End Open-Source Linux-Capable RISC-V SoC in 130nm CMOS
}

\begin{document}


\author{
\IEEEauthorblockN{%
Paul Scheffler\orcidlink{0000-0003-4230-1381}\IEEEauthorrefmark{1}\IEEEauthorrefmark{4}, %
Philippe Sauter\orcidlink{0009-0001-6504-8086}\IEEEauthorrefmark{1}\IEEEauthorrefmark{4}, %
Thomas Benz\orcidlink{0000-0002-0326-9676}\IEEEauthorrefmark{1}\IEEEauthorrefmark{4}, %
Frank K. Gürkaynak\orcidlink{0000-0002-8476-554X}\IEEEauthorrefmark{1}, %
Luca Benini\orcidlink{0000-0001-8068-3806}\IEEEauthorrefmark{1}\IEEEauthorrefmark{2}%
}
\thanks{%
    \IEEEauthorrefmark{4} Authors contributed equally to this research.
}
\IEEEauthorblockA{
    \textasteriskcentered~\textit{Integrated Systems Laboratory, ETH Zurich}, Switzerland \\
    \textdagger~\textit{Department of Electrical, Electronic, and Information Engineering, University of Bologna}, Italy \\
    \{paulsc,phsauter,tbenz,kgf,lbenini\}@iis.ee.ethz.ch
    }
}

\maketitle

\begin{abstract}
Open-source hardware (OSHW) is rapidly gaining traction in academia and industry.
The availability of open RTL descriptions, EDA tools, and even PDKs 
enables a fully auditable supply chain for end-to-end (RTL to layout) open-source silicon, significantly strengthening security and transparency. 
%
Despite promising developments, existing OSHW efforts have so far fallen short of producing end-to-end open-source SoCs at the complexity and performance level needed to run a general-purpose OS.
%
We present Basilisk, the first end-to-end open-source, Linux-capable RISC-V SoC taped out in IHP's open \SI{130}{\nano\meter} technology.
Basilisk features a 64-bit RISC-V core, a fully digital HyperRAM DRAM controller, and a rich set of IO peripherals including USB 1.1 and VGA.
To tape out Basilisk, we create a reusable tool pipeline to convert its industry-grade SystemVerilog description to Verilog.
We optimized logic synthesis in the open source Yosys synthesis tool, obtaining an increase in Basilisk's peak clock speed by \SI{2.3}{\x} to \SI{77}{\mega\hertz} and reducing its cell area by \SI{1.6}{\x} to \SI{1.1}{\mega\gateeq} while also reducing synthesis runtime and RAM usage.
We further optimized place and route in OpenROAD, enabling convergence to zero DRC violations while increasing core area utilization by \SI{10}{\percent} and reducing die area by \camr{\SI{12}{\percent}}.

\end{abstract}

\begin{IEEEkeywords}
Open-source hardware, SoCs, EDA, RISC-V
\end{IEEEkeywords}


\section{Introduction}

In recent years, \gls{oshw} has gained attention from industry and academia alike.
The increased momentum on open source hardware has led to the creation of open \gls{rtl} descriptions for \gls{ip} blocks~\cite{pulp-platform, 9097398, chen2020xuantie}, \gls{eda} tools~\cite{ajayi2019openroad, wolf2013yosys}, and even \glspl{pdk}~\cite{ihp-git}.
\gls{oshw} opens up a traditionally closed design process, curtailing or even eliminating \gls{ip} and tool licensing costs and enabling open research and collaboration without \glspl{nda}.

Most notably, \gls{oshw} enables a \emph{transparent} and \emph{verifiable} hardware supply chain from \gls{rtl} description to layout.
By combining open \glspl{ip}, \gls{eda} tools, and \glspl{pdk} into an end-to-end open-source flow, open hardware designers can empower third parties to not only reproduce the final layout, but to fully audit its design process and verify its logic equivalence to the original \gls{rtl} description. 
Instead of having to trust closed-source tools and \glspl{ip} from commercial providers, the functionality and security of \gls{oshw} can be verified independently across levels of abstraction.

Strengthening security through auditable \gls{oshw} is not a new idea; the \emph{OpenTitan} project~\cite{ot-git} provides open-source \gls{rot} \glspl{ip} and recently taped out a first chip with \gls{rot} functionality. 
However, while some OpenTitan test chips were designed using partially open tools, none of their existing designs are end-to-end open-source.
Moreover, even if applications can trust an open \gls{rot} with isolating security-critical data and operations, the remaining \gls{soc} is usually closed-source and cannot be trusted by programmers or end users.
Some partially open-source processors~\cite{zaruba2019ariane, chen2020xuantie} and \glspl{soc}~\cite{9097398, ottaviano2023cheshire} are available, but to date, there is no existing end-to-end open-source Linux-capable \gls{soc}.

In this work, we present Basilisk\footnote{\url{https://github.com/pulp-platform/cheshire-ihp130-o}}, the first end-to-end open-source, Linux-capable RISC-V \gls{soc} taped out in IHP's open \SI{130}{\nano\meter} technology.
Basilisk is based on the configurable Cheshire~\cite{ottaviano2023cheshire} \gls{soc} platform and combines an RV64GC core with a fully digital HyperRAM DRAM controller and a rich set of peripherals, including VGA and USB 1.1, to complete a useful real-world Linux system. 
Like its hardware, Basilisk's boot code and firmware are completely open; from reset to Linux init, all executed code is open-source and auditable.

To tape out Basilisk with competitive \gls{qor}, we vastly improve the state-of-the-art open \gls{eda} flow using \emph{Yosys}~\cite{wolf2013yosys} and \emph{OpenROAD}~\cite{ajayi2019openroad}.
First, we create a reusable tool pipeline simplifying Basilisk's industry-grade \gls{sv} \gls{rtl} description to Yosys-supported Verilog by introducing our parameter-resolving \gls{sv} pre-elaborator \emph{SVase}.
Then, we optimize Yosys' logic synthesis by improving multiplexer handling, integrating \gls{lms}~\cite{lazy-synthesis}, and mapping arithmetic units to a library of preoptimized designs. 
Finally, we improve the OpenROAD \gls{pnr} tool flow by designing a routing-friendly power grid and tuning global hyperparameters.
Overall, our flow optimizations improve Basilisk's peak clock frequency from \SI{33}{\mega\hertz} to \SI{77}{\mega\hertz}, reduce logic area from \SI{1.8}{\mega\gateeq} to \SI{1.1}{\mega\gateeq}, increase core utilization from \SI{50}{\percent} to \SI{55}{\percent}, and reduce synthesis runtime and peak RAM usage by \SI{2.5}{\x} and \SI{2.9}{\x}, respectively.

To summarize, our contributions are as follows:

\begin{itemize}

    \item We present Basilisk's open-source, extensible architecture featuring a Linux-capable 64-bit RISC-V core, a HyperRAM DRAM controller, a hierarchical interconnect, and a rich set of IO peripherals including  USB 1.1 and VGA.
    
    \item We create a reusable open-source tool pipeline simplifying Basilisk's industry-grade \gls{sv} \gls{rtl} description to a single Yosys-readable Verilog file by leveraging our parameter-resolving \gls{sv} pre-elaborator \emph{SVase}.
    
    \item We optimize Yosys' logic synthesis \gls{qor} by improving multiplexer handling, integrating \gls{lms}, and leveraging a library of optimized arithmetic units, increasing Basilisk's clock speed by \SI{2.3}{\x} and reducing its cell area by \SI{1.6}{\x}  while also reducing synthesis runtime and RAM usage.
    
    \item We improve the OpenROAD \gls{pnr} tool flow by designing a routing-friendly power grid and tuning global hyperparameters, 
    achieving zero design rule violations, improving core utilization by \SI{10}{\percent}, and reducing die area by \SI{12}{\percent}.
    
\end{itemize}

\section{Architecture}

\begin{figure}[t]
  \centering
  \includegraphics[width=\linewidth]{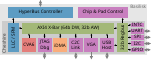}
  \caption{Top-level block diagram of Basilisk.}    
  \label{fig:arch} 
\end{figure}

Basilisk is built around \emph{OpenHWGroup}'s energy-efficient RV64GC CVA6~\cite{zaruba2019ariane} processor.
It is based on our open-source, silicon-proven \emph{Cheshire} \gls{soc} platform~\cite{ottaviano2023cheshire} designed to provide a minimal, highly configurable, autonomously booting 64-bit \riscv~host for Linux-capable systems.

\cref{fig:arch} shows Basilisk's top-level architecture.
It includes all hardware components necessary to boot and run Linux without support from an external host, such as RISC-V-compliant interrupt controllers, various standard IO interfaces, and a fully digital DRAM interface.
To balance performance and design complexity, Basilisk uses a two-stage interconnect: request initiators and high-throughput endpoints attach to a fully connected 64-bit \gls{axi}~\cite{9522037} crossbar, while low-throughput peripherals and configuration interfaces without burst support are accessed through a lightweight Regbus~\cite{regbus-git} demultiplexer.

Basilisk features a fully digital HyperRAM DRAM controller supporting two chips and a transfer speeds of up to \SI{154}{\mega\byte\per\second}. 
The controller is connected to the \gls{axi} crossbar through a 4-way, \SI{64}{\kibi\byte} \gls{llc};
each way can dynamically be configured as scratchpad memory to provide %
on-chip SRAM when needed.
CVA6 is configured with 2-way, \SI{16}{\kibi\byte} L1 instruction and data caches.%

Basilisk provides a rich set of peripherals.
In addition to I2C, quad SPI, and UART for serial communication, it includes a four-port USB 1.1 (OHCI) host controller and a VGA controller for video output.
\camr{Unlike later protocol revisions, USB 1.1 can be implemented using only digital logic and regular-speed IOs, keeping Basilisk's design highly accessible.}
Each USB port is multiplexed with GPIOs, providing a software-controlled IO bus up to 8 bit wide.
A JTAG test access point connected to a \riscv~debug module enables live debugging of the CVA6 core and full memory bus access.
A fully-digital\camr{,} double-data-rate, \camr{duplex \SI{77}{\mega\bit\per\second}} \gls{c2c} link serializing the \gls{axi} protocol allows two Basilisk chips to communicate \camr{through direct} interconnect accesses.
A high-efficiency, asynchronous DMA engine~\cite{10311078} capable of 2D transfers relieves CVA6 of data movement tasks.
All of Basilisk's peripherals are designed to be Linux-compatible, with many already having working drivers for our version of CVA6 Linux \camr{(kernel version 5.10.7, updates ongoing)}.

Basilisk's boot ROM enables autonomous boot from a \glsunset{gpt}\gls{gpt}-formatted SD card, SPI NOR flash, or I2C EEPROM.
It loads a small binary of up to \SI{48}{\kibi\byte} into its internal scratchpad, which in turn may load a firmware (e.g. OpenSBI) and a full-fledged bootloader (e.g. U-Boot) into DRAM.
Alternatively, code may also be preloaded through JTAG, UART, or the \gls{c2c} link.
In our upstream setup, all software run from SoC reset to Linux userspace is completely open-source and auditable, including the boot ROM, firmware, and our Linux modifications.

Basilisk is highly extensible and reconfigurable by design.
Adding interconnect ports and interrupts for new \glspl{ip}, removing existing blocks, or even using multiple CVA6 cores is simply a matter of reparameterization. 
We hope this will allow other designers to build on and extend Basilisk with minimal effort.

\section{Implementation Flow}

We implement Basilisk in IHP's open \SI{130}{\nano\meter} technology using \emph{Yosys} and \emph{OpenROAD}. 
To this end, we
implement a reusable tool pipeline simplifying Basilisk's industry-grade \sv~\gls{rtl} description to Verilog supported by Yosys (\cref{sec:impl_rtlpre}), %
enhance synthesis to significantly improve \gls{qor} while reducing runtime and memory footprint (\cref{sec:impl_synth}),
and optimize the reference backend flow to improve \gls{qor} and minimize design rule violations (\cref{sec:impl_pnr}).

\subsection {RTL Description Preprocessing}
\label{sec:impl_rtlpre}

Yosys currently cannot read in synthesizable \sv, supporting only Verilog-2005 and a few selected \sv~constructs.
Existing open-source solutions to synthesize \sv~designs in Yosys include the \sv-to-Verilog conversion tool \emph{SV2V}~\cite{sv2v-git} and the third-party \emph{Synlig}~\cite{synling-git} frontend.
Unfortunately, these solutions cannot handle Basilisk's industry-grade \gls{rtl} description; both fail to correctly resolve hierarchically propagated design parameters%
, which are essential to keeping complex, parametric designs manageable without resorting to code generation.

Prior to our work, the only open-source tool correctly resolving Basilisk's parameterization was \emph{Slang}~\cite{slang-git}, a library providing full \gls{sv} elaboration and the best \gls{sv} language support of all open-source tools evaluated by ChipsAlliance's \gls{sv} test suite~\cite{sv-tests-git}.
However, as a library, Slang only provides elaborated designs as in-memory data structures with no existing solution to pass them on to Yosys for synthesis.

To close this flow gap, we created \emph{SVase}~\cite{svase-git}, a source-to-source \gls{sv} pre-elaborator leveraging Slang. 
SVase rewrites all parameter expressions as literals, unrolls all generate constructs, and uniquifies every used module parameterization.
The resulting \gls{sv} \gls{rtl} description has \emph{no} unresolved parameters or dependencies between instances and is simple enough to correctly be translated to Verilog by SV2V.

\begin{figure}[t]
  \centering
  \includegraphics[width=\linewidth]{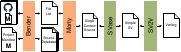}
  \caption{Basilisk's SV-to-Verilog RTL description preprocessing flow.}    
  \label{fig:impl_rtlpre} 
\end{figure}

For simplicity, SVase assumes that all input \gls{sv} sources are collected in a single file, which we automate using our existing source management tool \emph{Bender}~\cite{bender-git} and our \gls{sv} source pickler \emph{Morty}~\cite{morty-git}.
\cref{fig:impl_rtlpre} shows the resulting end-to-end tool flow converting Basilisk's multi-file \gls{sv} \gls{rtl} description to a single Verilog file readable by Yosys.
We emphasize that this flow is not specific to Basilisk or Yosys; it can simplify \emph{any} industry-grade \gls{sv} design to a single Verilog file for use with \emph{any} tool, including simulators with limited \gls{sv} support.
\camr{Furthermore, the full flow takes less than \emph{two minutes} to run on Basilisk, which is two orders of magnitude less than Yosys synthesis after our optimizations (\SI{2.2}{\hour}) and comparable to elaboration steps in commercial, \gls{sv}-capable simulators and synthesis tools.}

\subsection {Synthesis Enhancements}
\label{sec:impl_synth}

Before our work, synthesizing Basilisk with Yosys resulted in inacceptable \gls{qor} and inflated runtime and memory usage (see \cref{sec:impl_pnr}). 
We present three key, design-independent improvements to Yosys' logic synthesis that drastically improve design \gls{qor} and Yosys' resource footprint.

\subsubsection*{Part-select synthesis}

Yosys versions prior to our improvements (\textless 0.34) use generic shift operations (\texttt{\$shiftx}) to represent all indexed part-select operations instead of more efficient \emph{block-multiplexer trees} where possible.
Thus, for any part select, a generic barrel shifter supporting an arbitrary shift amount is inferred at elaboration.
This broad generalization significantly inflates area %
and increases runtime and peak memory usage; it is especially problematic in conjunction with SV2V, which translates all selections into arrays of packed data as part selects.
Later logic optimization stages are unable to simplify these shifters to the desired multiplexer trees, strongly impacting \gls{qor}. 
Instead of changing the representation of part-selects, we develop a new optimization pass that identifies shift operations with constant strides, pads the implied blocks to a power of two, and thus enables existing optimizations to remove unnecessary logic.
Compared to a solution inferring block multiplexers, our approach achieves the same results on part selects, but overall superior results as other shift operations may also benefit from this optimization.

\subsubsection*{Lazy man's synthesis}

In cooperation with logic synthesis researchers and ABC developers, we overhaul the ABC script, leveraging Yang et al.'s work on \gls{lms}~\cite{lazy-synthesis} to improve \gls{qor} at the cost of minimal additional runtime.
Near-optimal implementations of six-input, one-output logic functions are precomputed using other optimizers in ABC. 
The functions and implementations, together with their characteristics, are stored in a look-up table. 
This process is time- and resource-intensive, but only needs to be performed once, and the resulting table can be re-used across different designs.
During synthesis, the netlist is divided into blocks with six inputs and one output and the table is probed for the best fitting implementation, replacing each block with the corresponding structure.

\subsubsection*{Library of Arithmetic Units}

Yosys currently uses a suboptimal approach to implement fundamental arithmetic units.
Addition operations infer one globally selected adder architecture, impeding a balanced area-speed tradeoff.
Multiplication operations are implemented using Booth's algorithm.
More complex operations are implemented using the basic operation mappings; in the case of the \gls{mac} operation, a Booth multiplier followed by an adder is inferred. 
A more efficient solution is to integrate adders into the \emph{CSA} tree of preceding multipliers, creating \gls{fma} units.
We use our library of optimized arithmetic unit implementations to map a timing-critical \gls{mac} operation as an \gls{fma} unit, shortening our critical path.
We further replace the default adder implementation in Yosys with a selection of optimized adder architectures.
A solution to automatically infer \glspl{fma} and other fused operations and implement them from our library is in the works.

\subsection {Place and Route Optimizations}
\label{sec:impl_pnr}

We use \emph{OpenRoad}~\cite{ajayi2019openroad} to place and route Basilisk's synthesized netlist. 
We mainly identify possible improvements in the \emph{\gls{eda} tool flow} (how the individual components of OpenROAD are invoked) and the \emph{physical constraints} of the \gls{asic}.
%
We improve the routability of the design by redesigning the power grid;
we reduce the width and increase the count of the power stripes on the top metal layer to ease 
routing congestion underneath the stripes. 
%
Very dense modules with random routing patterns, such as the boot ROM, were a particular source of issues. As OpenRoad currently only accepts global (as opposed to region- or instance-based) settings, we tune several \emph{hyper-parameters} of the routability-driven global placement engine to improve the placement of dense blocks and get a routable design without \gls{drc} violations.

\section{Results}

We present the final \gls{qor} of Basilisk and quantify our synthesis and \gls{pnr} optimizations in relation to a baseline Yosys-and-OpenROAD flow without improvements.

\subsection{Synthesis}

\begin{figure}[t]
  \centering
  \includegraphics[width=\linewidth]{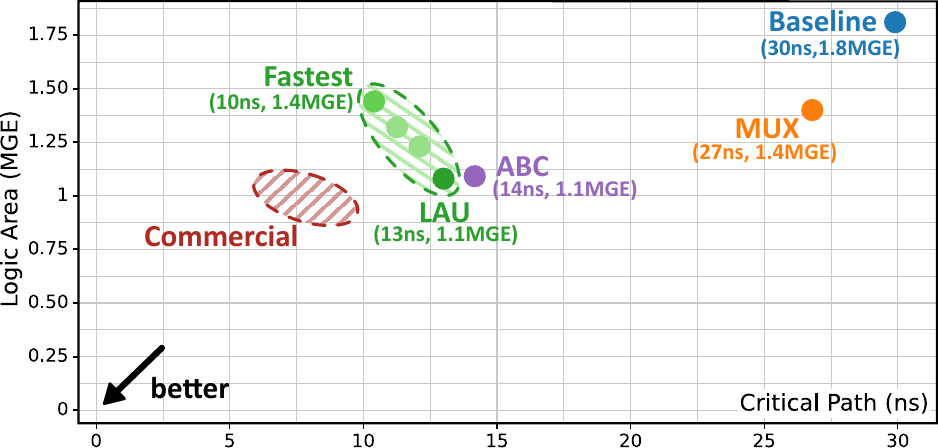}
  \caption{Area-time plot summarizing the incremental \gls{qor} benefits of our Yosys synthesis optimizations and comparing them to commercial \gls{qor}.}    
  \label{fig:at-plot} 
\end{figure}

\Cref{fig:at-plot} summarizes the cumulative \gls{qor} effects of our synthesis enhancements in an \gls{at} plot.
We time our netlists in a commercial tool to ensure accurate results under typical conditions (\SI{1.2}{\volt}, \SI{25}{\celsius}).
We compare our work to a \emph{baseline} Yosys flow without our optimizations using the default, speed-optimized ABC script from the OpenROAD flow scripts; it yields a logic area of \SI{1.8}{\mega\gateeq} and a critical path length of \SI{30}{\nano\second} (\SI{33}{\mega\hertz}). 

Our first optimization improving the synthesis of part-selects (\emph{MUX}) reduces logic area by \SI{22}{\percent} and the critical path length by \SI{11}{\percent}.
Building an optimized ABC script that leverages \gls{lms}  (\emph{ABC}) yields the largest \gls{qor} improvement, further reducing area \camr{by} \SI{21}{\percent} and shortening the critical path by another \SI{1.9}{\x}.
Finally, using our optimized library of arithmetic units (\emph{LAU}) further shortens the critical path by \SI{9}{\percent} to \SI{13}{\nano\second} (\SI{77}{\mega\hertz}). 
Together, our Yosys optimizations reduce synthesis time from \SI{5.4}{\hour} to \SI{2.2}{\hour} (\SI{2.5}{\x}) and peak synthesis RAM usage from \SI{217}{\giga\byte} to \SI{75}{\giga\byte} (\SI{2.9}{\x}).

Our parametric {ABC} script and the span of choices in our library of arithmetic units can provide designers with flexible control over area-timing tradeoffs, allowing us to generate a pareto-frontier of multiple designs as shown.
While the minimal-area design we ultimately chose (\SI{1.1}{\mega\gateeq}, \SI{13}{\nano\second}) improves timing by \SI{2.3}{\x} and reduces the area by \SI{1.6}{\x}, tuning our ABC script for timing further reduces the critical path to \SI{10.4}{\ns} (\SI{97}{\mega\hertz}) at the cost of notably increased area.

Despite our significant \gls{qor} improvements, commercial synthesisis tools still have a clear edge on multi-million-gate designs like Basilisk. 
They achieve their superior \gls{qor} through timing-aware synthesis, tighter integration of elaboration and optimizations, larger libraries of pre-optimized blocks, and a stronger emphasis on physically aware synthesis.
Nevertheless, our optimizations take a significant step toward closing the \gls{qor} gap between open-source synthesis and commercial flows; our best logic area and critical path length are both within \SI{50}{\percent} of what a commercial synthesis tool achieves.

\subsection{Place and Route}
\begin{figure}[t]
    \centering
    \subfloat[\label{fig:die:iguana}]{{\includegraphics[width=0.455\linewidth]{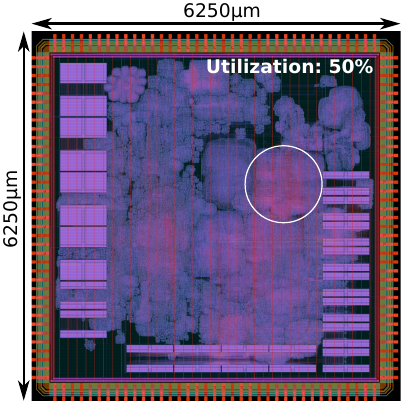}}}%
    \hspace{0.1cm}
    \subfloat[\label{fig:die:basilisk}]{{\includegraphics[width=0.455\linewidth]{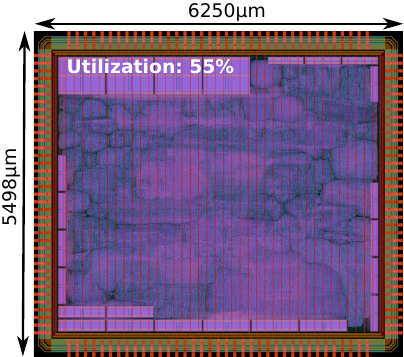}}}%
    \caption{Basilisk layouts produced by the baseline (a) and optimized flow (b). The white circle highlights excessive top metal routing (red) in the baseline.}    
    \label{fig:die-shots} 
\end{figure}

\Cref{fig:die-shots} shows the final Basilisk layout from OpenROAD without and with our synthesis and \gls{pnr} \gls{qor} optimizations.
In the former (baseline) case, we re-synthesize some problematic modules including the boot ROM, the CVA6 issue stage, and the L1 data cache with a commercial tool to avoid an unmanageable number of \gls{drc} violations \camr{ primarily caused by inefficient part-select handling in synthesis (See \cref{sec:impl_synth})}. 

Our improvements to the physical implementation flow increase core area utilization from \SI{50}{\percent} to \SI{55}{\percent} (+\SI{10}{\percent}) while reducing local peak routing resource utilization, enabling OpenROAD to converge to \emph{zero} \gls{drc} violations. 
When using only Yosys for baseline synthesis, we reduce the peak routing resource utilization, found in the boot ROM, from an unfeasible \SI{210}{\percent} to \SI{100}{\percent}.
Tuning OpenROAD's hyperparameters reduces routing congestion both on the global and local scale, obviating most signal routing on the unsuited top metal layers and resulting in a less clustered floorplan.
Overall, we reduce Basilisk's die area from \SI{39}{\milli\meter^2} to \SI{34}{\milli\meter^2} (-\SI{12}{\percent}).

\section{Conclusion and Outlook}

We present Basilisk, \camr{the} first end-to-end open-source, Linux-capable RISC-V SoC taped out in IHP's open \SI{130}{\nano\meter} technology.
Basilisk features a 64-bit RISC-V core, a fully digital HyperRAM DRAM controller, and a rich set of IO peripherals including USB 1.1 and VGA.
To tape out Basilisk, we create a reusable tool pipeline converting its industry-grade SystemVerilog description to Yosys-readable Verilog.
We also significantly optimize Yosys' synthesis \gls{qor}, improving Basilisk's clock speed by \SI{2.3}{\x} to \SI{77}{\mega\hertz} and reducing cell area by \SI{1.6}{\x} to \SI{1.1}{\mega\gateeq} while also reducing synthesis runtime and peak RAM usage by \SI{2.5}{\x} and \SI{2.9}{\x}, respectively.
Finally, our OpenROAD \gls{pnr} optimizations enable convergence to zero \gls{drc} violations, improve core area utilization by \SI{10}{\percent}, and reduce die area by \camr{\SI{12}{\percent}}.
In future work, we hope to enhance Basilisk with open-source \gls{rot} \glspl{ip} to also provide robust cryptographic security and a verified boot chain.

\section*{Acknowledgement}
We thank %
A. Mishchenko, %
M. Fujita, %
G. Micheli, %
A. Costamagna, %
A. T. Calvino, %
O. Hammad, %
M. Liberty, %
M. Povišer, %
the Yosys team, %
B. Muheim, %
and %
Z. Jiang, %
for their valuable contributions to the research project. %
We further thank all contributors to the open-source EDA tools. %
We are deeply grateful to IHP for their generous support and providing us with the opportunity for an open-source tapeout at this scale.

\bibliographystyle{IEEEtran}
\bibliography{main.bib}

\end{document}